# Interatomic exchange coupling of BCC iron

Hai Wang, Pui-Wai Ma and C. H. Woo[*]

*Department of Electronic and Information Engineering, The Hong Kong Polytechnic University,*

*Hong Kong SAR, China*

**Abstract**

We performed first-principle calculations on the exchange interaction (EI) between atoms in BCC-Fe strained volumetrically. Our results show that the volume-dependence of the EI deviates considerably from the Bethe-Slater curve. This behavior is discussed in terms of the on-site and/or inter-site direct exchange interactions between electrons.

PACS number(s): 75.30.Et, 71.15.Mb, 75.50.Bb, 71.20.Be

75.30.Et Exchange and superexchange interactions (see also 71.70.Gm Exchange interactions)

71.15.Mb Density functional theory, local density approximation

75.50.Bb Fe and its alloys

71.20.-b Electron density of states and band structure of crystalline solids

71.20.Be Transition metals and alloys

---

[*] corresponding author, email address: chung.woo@polyu.edu.hk

**I. Introduction**

Magnetic effect is well known to play a pivotal role in the structural stability of iron [1]. For example, the softening of tetragonal shear modulus $C' = \frac{1}{2}(C_{11} - C_{12})$ at elevated temperature [2] [3] [4] and the resulting change of dominant orientation of dislocation loops from <110> to <100> at around 500°C, are believed to be results of the magnetic effect [5] [6] [7]. Lavrentiev *et al.* [8, 9] also showed that magnetic fluctuations are responsible for the BCC-FCC phase transition of iron at 1184K, which is just slightly higher than the Cure temperature $T_C$ at 1043K.

Although a*b initio* electronic calculations are powerful tools in the exploration of ground-state properties of materials, it is in general difficult to extend its application to practical cases where the involvement of heat and thermodynamics and effects of the many-body elementary excitations are important. Attempts to use the Fermi distribution to mimic finite temperature behavior of electrons in materials [10] are restricted to cases where the dynamic interactions among the electrons, the spins and the lattice are sufficiently weak, which almost certainly excludes the strong-interaction regime near phase transitions.

In order to investigate the strongly interactive spin and lattice dynamics at higher temperatures, Antropov *et al.* [11] suggested that the interactions between atoms, including the interatomic exchange coupling, can be obtained via *ab initio* calculations. The dynamics of the interactive atoms and the coupled spins, as represented by the Heisenberg Hamiltonian, can then be integrated using molecular dynamics (MD) and classical spin dynamics (SD), respectively. Nevertheless, the suggestion remains untested. The computer resources required are expected to be large and could be prohibitive in the actual implementation.

By explicitly incorporating a magnetic term in the functional form of the empirical many-body potential, Dudarev and Derlet [6-8] proposed the first magnetic potential for MD simulation at 0K. The magnetic effect due to this potential stabilizes the dumbbell configuration so that the ground-state self-interstitial configuration in BCC iron is not the <111> crowdion but the the <110> dumbbell. This result is consistent with the *ab initio* calculations [12].

Recently, Ma *et al*. [13] [14] successfully developed the spin-lattice dynamics (SLD)

simulation scheme which incorporates both the spin and lattice degree of freedom into a set of coupled equations of motion. The dynamics of the coupled spin and lattice subsystems can be described in terms of the empirical many-body potential $U(\mathbf{R})$ and the exchange coupling function $J_{ij}(\mathbf{R})$ between atoms $i$ and $j$ in the Heisenberg Hamiltonian. Here $\mathbf{R}=\{\mathbf{R_i}\}$ stands for the set of atomic position vectors and represents the atomic configuration. The interaction between spin and lattice vibrations in SLD is considered through the configuration dependent exchange coupling function $J_{ij}$. Ma *et al.* [15] demonstrated for BCC iron thin film the importance of considering the coupling between the lattice and the magnetic vibrations (i.e., the phonon-magnon interaction), which quantitatively and qualitatively alters the mechanical responses.

As an initial attempt, Ma *et al.* [15] assumed a pairwise form for $J_{ij}$ neglecting many-body effects. Although this approach appears to yield reasonable results in the study of perfect crystal properties, its applicability may be questionable in studies involving lattice defects, such as dislocations, interstitial clusters, high-energy displacement cascades, etc., where near-neighbor atomic configurations may deviate significantly from those of the perfect crystal. Compounding the problem, the volume of data in the literature useful for the determination of $J_{ij}$ is scarce, except for the limited regime near the equilibrium configuration [16] [17] [18] [19]. There is a drastic lack of information for the consideration of $J_{ij}$ for the more general applications. For example, although the magnetic moment (MM) of iron is well known to diminish with increasing atomic density, vanishing beyond a sufficiently high atomic density [20] [21], yet information regarding the details of this behavior is not available, making the estimation of $J_{ij}$ in this regime even more difficult.

In this paper, we perform ab initio calculations on $J_{ij}$ between different neighbors of BCC-Fe for various volume strains. The results are used to study the behavior of $J_{ij}$ and how it varies with the configuration of the neighboring atoms. We use spin polarized scalar-relativistic linear muffin-tin orbitals (LMTO) method combined with the Green's function (GF) technique [22]. The accuracy of the electronic structure is cross checked with WIEN2k [23]. Many-body effects are found to be very significant in the magnitude of $J_{ij}$. A pairwise functional representation of $J_{ij}$ independent of the neighboring environment is found to be inadequate. For the nearest neighbors, $J_{ij}$ generally decreases monotonically with increasing separation $R_{ij}$, except for an unexpected local minimum when $R_{ij}$ is slightly larger than the equilibrium distance. The magnitude of the MM also shows a sudden increase in the same regime. The values of $J_{ij}$ between second nearest neighbors are in general substantially smaller than the first nearest neighbors, except for distances larger than the equilibrium lattice distances. At smaller distances, on the other hand, the values of $J_2$ decreases, rather than increases, with decreasing $R_{ij}$ even turn negative for distances sufficiently small. These results will be discussed and its implications analyzed.

## II. Methodology

Only BCC iron is considered in this paper. For lattice constants ranging from 2.2Å to 3.2Å, we calculate the magnitude of atomic magnetic moment, densities of state, charge density maps and the exchange coupling constant $J_{ij}$. We are aware that changing the atomic volume may induce phase transition [24] [25], nevertheless, this is beyond our present scope. Two computer codes are employed, WIEN2k [23], which is based on the full-potential linearized augmented plane-wave (FP-LAPW) method, and LMTO-GF [26], which is implemented within the atomic spheres approximation (ASA) [22]. While WEIN2k may be superior for electronic structure calculations, it is not equipped to calculation $J_{ij}$, as LMTO-GF is. Both computer codes, however, can be used to calculate MM.

The $Im\bar{3}m$ space group is used for BCC iron. Brillouin-zone integrals are calculated using a 40x40x40 $k$-points mesh. There are all together 1661 irreducible $k$-points. For the generalized gradient approximation (GGA), the Perdew-Burke-Ernzerhof (PBE) scheme is used [27]. For the local density approximation (LDA), Perdew and Zunger (PZ) [28] is used in WIEN2k, and von Barth and Hedin (vBH) [29] is used in LMTO-GF. Since GGA is better than LDA in describing the ground-state structure of iron [30], most investigations on the electronic and magnetic properties of BCC iron are carried out by using GGA [31]. In the following calculations, we mainly use the GGA. However, LDA is also used whenever the comparison is meaningful.

We set $R_{mt}K_{max}=10$ in the WIEN2k calculation, where $R_{mt}$ is the muffin-tin radius set (manually) to maximize the muffin-tin sphere volume, and $K_{max}$ the maximum reciprocal space vector. We also set the maximum number of the spherical harmonics for the expansion of waves inside the atomic spheres, i.e., $l_{max}+1$, to 12, and the largest reciprocal vector in the charge Fourier expansion $G_{max}$ to 15. In LMTO-GF, where the Green function calculation is carried out, $R_{mt}$ is set equal to the radius of the Wigner-Seitz cell.

LMTO-GF calculates the values of $j_{ij}$ in the effective Hamiltonian [32]

$$H = -\sum_{i \neq j} j_{ij} \mathbf{e}_i \cdot \mathbf{e}_j \qquad (1)$$

within ASA, by converting the scattering operators to Green's functions [33] using the linear response theory. In equation (1), $\mathbf{e}_i$ is the unit vector in the direction of the MM at site $i$. The summand in the Hamiltonian expresses the magnetic energy between the pair of atoms $i$ and $j$. We note that $j_{ij}$ defined in eq. (1) differs from the exchange coupling parameter $J_{ij}$ in the Heisenberg Hamiltonian, but is related to it by $J_{ij} = j_{ij}/\ ^2$ where is the total MM per unit atomic volume. We also note that $j_{ij}$ varies in different atomic configurations, due to the modification in the overlap of atomic orbitals [34].

Using the Andersen magnetic force theorem [35], $j_{ij}$ can be rewritten [26] as

$$j_{ij} = \frac{1}{2\pi} \int^{\varepsilon_F} d\varepsilon \, \mathrm{Im} Tr_L \left\{ \delta P_i \left( T_{ij}^{\uparrow} T_{ji}^{\downarrow} + T_{ij}^{\downarrow} T_{ji}^{\uparrow} \right) \delta P_j \right\} \qquad (2)$$

where $T_{ij}^{\uparrow/\downarrow}$ is the scattering operator and $\delta P_i$ the on-site perturbation. Once we obtain the $j_{ij}$, we may estimates the $T_C$ according to the mean-field approximation (MFA) [26]:

$$T_C^{MFA} = \frac{2}{3k_B} j_0 \qquad (3)$$

where $j_0 = \sum_{j \neq 0} j_{0j}$. Although the MFA is well known to overestimate the $T_C$ by about 1/3, we may still use it as an estimation for checking purposes.

The convergence of $j_{ij}$ with respect to the number of $k$-points is assessed with a progressively finer mesh. We use a mesh of 70x70x70 $k$-points corresponding to 8112 irreducible k-points. For a lattice constant of 2.8665Å, $j_{ij}$ converges with an accuracy of ~2% while the total energies converges to within ~0.5%. This is the tolerance we set for the present calculations.

**III. Results and Discussions**

Fig. 1 shows the total energy versus lattice constant we obtain with WIEN2k and LMTO-GF using GGA-PBE. The results from WIEN95 and the experiments are also shown for comparison. The agreement is within the tolerance of 0.5% of the calculation. Fitting the energy-volume data to Birch-Murnaghan equation of state [36], the equilibrium lattice constant, bulk modulus and pressure are obtained. Values of the lattice constant, and the corresponding MM, and bulk modulus at equilibrium calculated with WIEN2k are 2.873 Å, 2.25 $\mu_B$, and 1.73 Mbar, respectively. Analogous values obtained with LMTO-GF are 2.881 Å, 2.26 $\mu_B$, and 1.79 Mbar. All of them are in excellent agreement with each other and with the experimental values of 2.8665 Å [37], 2.22 $\mu_B$ [38], and 1.72 Mbar [38], and are consistent with the results of other theoretical works [1] [31] [39].

Using the same set of data, we also calculate the MM as a function of the lattice constant and plot the results in Fig. 2. Results calculated following Kormann *et al.* [40] are also shown for comparison. Good consistency can be seen among results from the different methods. Except for small lattice constants, the MM calculated with WIEN2k and LMTO-GF vanish at around 2.3 Å to 2.4 Å, showing a magnetic/non-magnetic (i.e., Stoner) transition at high atomic density [25] [41]. The difference between the results of the two methods is mainly due to the use of the ASA in the LMTO-GF model. Indeed, since the relative geometric difference between the Wigner-Seitz cell and an atomic sphere of equal volume increases with decreasing atomic volume, the error of the LMTO-GF model relative to the WIEN2k model also tends to becomes more significant, as can be seen from Fig. 2. Nevertheless, the overall agreement between the MM from the two models remains good, showing the magnetic/non-magnetic transition at a lattice constant of ~2.3 Å, and the characteristic inflection point near ~2.90 Å. The present results are consistent with those obtained by Kormann *et al.* [40] using VASP with GGA, which is also shown in the inset of Fig. 2. The characteristic inflection point of the MM also occurs in our calculations when LDA is used, but not in Ref. [40], where it appears at a slightly larger lattice constant. There is no clear explanation of this characteristic inflection point of the MM in the literature.

The partial densities of states (DOS) for lattice constants 2.25 Å, 2.45 Å and 2.88 Å

calculated using LMTO-GF and WIEN2k are plotted in Figs. 3a-b. Good agreements are found between the two models in all cases. Proceeding from (i) to (ii) of Figs.3a-b, pressure reduction, corresponding to an increase of the lattice constant from 2.25 Å to 2.45 Å, removes the spin-degeneracy, resulting in the splitting of the density of state (DOS) and the restoration of the ferromagnetic state. In (iii) of Figs.3a-b, the uncomplete filled majority-spin subband at the equilibrium lattice constant 2.88 Å is consistent with the weakly ferromagnetic nature of BCC iron. In these figures, we also note the small presence of the *s*- and *p*-contributions to the total DOS, which is practically completely dominated by the *3d* contributions. The foregoing results are consistent with the findings of [42, 43] that the importance of Coulomb correlations decreases under pressure, and that ferromagnetic metals like Fe, Co, and Ni are all presumed non-magnetic metals under high pressure [24].

BCC iron has $O_h$ symmetry and the *3d*-DOS is split into the $t_{2g}$ (triply degenerate) and $e_g$ (doubly degenerate) components. The partial DOS of the $t_{2g}$ and $e_g$ states are shown in Figs. 3c-d. Each spin subband is labeled to show the occupancy number, obtained by integrating the partial DOS up to the Fermi level. Both the $t_{2g}$ and $e_g$ bands are about 60% filled in all cases. As the atoms in the highly compressed non-magnetic state move apart, the Pauli exclusion that favors the anti-parallel spin configuration weakens and the Coulomb exchange correlation that favors the parallel spin configuration starts to dominate. Splitting of both the $t_{2g}$ and $e_g$ spin subbands starts to occur beyond 2.3 Å (Fig. 2) as can be seen in (ii) of Fig. 3c-d. The splitting becomes very obvious in (iii) of Fig. 3c-d. Indeed, in this case, there are twice as many electrons with majority spin than with minority spin in the $t_{2g}$ band and that proportion is ~3.5 times in the $e_g$ band. The *3d* majority spin subband is almost, i.e., 80-90%, full. The minority spin subband, on the other hand, are only ~37% occupied and predominantly (~72%) by $t_{2g}$ electrons. This is consistent with the results of Jones et. al. [44]. Unequal occupancy of the spin subbands creates a MM and drives the non-magnetic/ferromagnetism transition as the lattice parameter increases beyond 2.3 Å.

The total DOS for the near-equilibrium structure, with lattice constants, 2.77 Å, 2.88 Å, 2.91 Å and 2.94 Å, are shown in Fig. 4a. Contributions from the $t_{2g}$ and $e_g$ bands are also respectively presented in Figs. 4b and 4c. In this range, one may notice that changing the lattice parameter has more influence on the band width, energy shift and the DOS at the Fermi level of the $t_{2g}$ state than the $e_g$ state. This is also reflected in the inflection point of the magnetic moment as a function of the lattice constant seen in Fig. 2, which is related to the relatively large change of the $t_{2g}$ DOS at the Fermi level due to the disappearance of $t_{2g}$ overlap (later we will shown).

In Fig. 5, we show the corresponding charge density maps of BCC-Fe on the (110) plane for the four lattice constants as in Fig. 4. The two columns are for the majority and minority spins as marked. The *x*-axis is along the <110> third nearest neighbors (3NNs) directions, the *y*-axis is along the <001> second nearest neighbors (2NNs) directions, and the diagonal is along the <111> nearest neighbors (1NNs) directions. Thus, AB and AC are 1NNs, BC and B'C' are 2NNs, and BB' and CC' are 3NNs. The topology of the electronic structure for the equilibrium lattice constant (2.88Å) presented in Fig. 5 is consistent with those obtained by Jones et. al. [44]. The difference between the electron distributions for two spin densities is obvious. Minority spins show stronger bonding tendency than the majority spins. While minority spins are mostly in the $t_{2g}$ states (note the square-shape of the electron cloud), the majority spins are more equitably distributed (note the circular-shape of the electron cloud) among the $t_{2g}$ and $e_g$ states. This is consistent with the corresponding partial DOSs shown in Figs. 4b-c, which we have discussed in the foregoing paragraph. The $t_{2g}$ band is due to the interatomic overlap between the $\{d_{xz}\}$ and $\{d_{yz}\}$ atomic orbitals of 1NNs in the <111> directions, while the $e_g$ band is between the $\{d_{z^2}\}$ atomic orbitals in the <001> directions. Consistent with the findings of Jones et. al. [44] the $e_g$ states in BCC Fe increases the charge density on the cube surfaces (see the charge clouds along BB'C'C in Fig. 5). The overlaps of the $e_g$ states may thus contribute to the spin correlation among both 1NNs and 2NNs, so that the exchange interaction between the 1NNs may have two contributions, one from the $t_{2g}$ state and one from the $e_g$ state, but those between the 2NNs only have contributions from the $e_g$ state.

As the interatomic separation increases, the electronic states become increasingly localized and the overlap between the atomic orbitals decreases. It can be seen from Fig. 5 that the overlapping of the $t_{2g}$ states between 1NNs decreases fast, and at a lattice constant of 2.94 Å, i.e., an interatomic separation of 2.55 Å, it is basically cut off for both majority (2.94

Å) and minority states (2.91 Å). This decrease is consistent with the rapid narrowing of the $t_{2g}$ band as the lattice constant increases in Fig. 3c and 4b. Overlap of the $e_g$ states, on the other hand, decreases slower, as can also be seen from the smaller change of the $e_g$ band as shown in Fig. 3d and 4c.

The spin densities, i.e., the number of electrons per atom $N_l$ with orbital angular momentum $l$ (= $s$, $p$, $d$), can be calculated by integrating the corresponding partial DOS up to Fermi level, from which the corresponding local MMs ( $N_l^\uparrow - N_l^\downarrow$) can also be obtained. The results are plotted as a function of the lattice constant in Fig. 6. In Fig. 6a, the majority spin density of $d$-electrons increases from 3.7 to 4.6 per atom while the minority decreases from 2.4 to 1.9 as the lattice constant increases from 2.45 Å to 2.96 Å. The corresponding MM in Fig. 6b is an increasing function of the lattice constant, which practically accounts for all of the contributions to the MM in Fig. 2. That the MM is mainly due to the $d$ orbitals and that the $s$, $p$ contributions are negligible can also be seen directly from their respective spin densities in Fig. 6b. The foregoing results are consistent with Frota-Pessoa *et al.* [17] who showed that the contributions to $j_{ij}$ is mainly due to the $d$-$d$ interaction (+110%), contributions from the $p$-$d$ (-7%), and the $s$-$d$ interaction are relative small.

Although it is generally accepted that GGA can give a better description of iron in many aspects [30], calculations of $j_{ij}$ in the literature are mostly performed with LDA. To adopt GGA for our calculations, we perform a comparison of the $j_{ij}$ obtained from both methods for the experimental lattice constant (2.8665Å). The results are listed in Table I, in which $j_n$ for $n$ = 1,2,3… represents the value of $j_{ij}$ between the $n^{th}$ nearest neighbors and $j_0$ is defined as in eq (3). Our LDA results are in good agreement with those of other authors [16] [17] [18]. Due to electron screening, except for the first two neighbor shells, $j_{ij}$ is at least an order of magnitude smaller. In this regard, we note that the accuracy of the smaller components is limited, with large percentage error as a result of the tolerance set to maximize the efficiency of the calculation. The results reported in the following are obtained only using GGA.

Now, we show in Fig. 7a the panoramic view of $j_n$, $n$ =1 to 8, as a function of lattice constant in rather large extent. For lattice constants smaller than 2.4 Å, the system is nonmagnetic and all $j_n$ vanishes. For all magnetic case, basically, $j_1$ and $j_2$ constitute the dominant contributions to the Heisenberg Hamiltonian. The LDA results of $j_1$ to $j_4$ obtained by Moran *et al.* [24] are also shown for comparison. Despite the much larger absolute values of their $j_1$ and $j_2$ compared with the those of [16] [17] [18] and with our present results, the slopes are very similar. In Fig. 7c, we plot against $R_{ij}$, the separation between atoms $i$ and $j$, the exchange coupling parameter $J_{ij}$ in the Heisenberg Hamiltonian from Figs. 6b and 7a, according to $J_{ij} = j_{ij}/\mu^2$ where $\mu$ is the total MM per unit atomic volume. We note that the dependence of $J_{ij}$ as a function of $R_{ij}$ shows more simple and linear than that of $j_{ij}$ in Fig. 7a. Fitting $J_2$ to linear function y(x)=a*x+b, we obtained that $a$=-1.63 and b=4.77 for $R_{ij}$ less than 2.94 Å, while $a$=0.07 and b=-0.15 for larger $R_{ij}$. Now, the dependence of $J_1$ is also more complex to fit to a linear function due to its observable two-peak character.

As discussed in the foregoing, analysis based on the DOS and the topology of the electronic distribution suggests that the exchange interaction between the 1NNs may have two contributions, one from the $t_{2g}$ state and one from the $e_g$ state, but those between the 2NNs only have contributions from the $e_g$ state. This picture is consistent with the double-peak structure of $j_1$, and the single-peak structure of $j_2$ in Fig. 6a, if we speculate that the two peaks of $j_1$ are from contributions $j_t$ from the $t_{2g}$ state and $j_e$ from the $e_g$ states, respectively, and the single peak of $j_2$ is from the $e_g$ contribution. In fact, we have previously shown the concrete evidence of the disappearance of $t_{2g}$ overlapping when lattice constant larger than 2.94 Å. Thus, by assuming that $J_t$ vanishes for $R_{ij} > 2.55$ Å, and that $J_e = J_2$ ($J_1$) for $R_{ij}$ less (larger) than 2.55 Å, $J_1$ can be written as a sum of $J_t$ an $J_e$, i.e., $J_1 = J_t + J_e$, This can be seen in Fig. 7d. This clearly confirms that the $J_{ij}$ are dominated by on-site and/or inter-site interaction, and the critical point at the ineratom distance of about 2.55 Å (lattice constant is 2.94 Å), less this both interactions are work while only the intersite one work when larger this. Through linear fitting, three sets of *a,b* values are obtained. It is exciting that these *ab inition* $J_{ij}$ data given in the form of linear function should be easy to introduce to SLD.

As the atoms in the highly compressed non-magnetic state move apart, the Pauli exclusion that favors anti-parallel spin configuration weakens and the Coulomb exchange starts to dominate and favor the parallel spin configuration. Accordingly, the exchange coefficient $J_{ij}(R_{ij})$ in the Heisenberg Hamiltonian would be negative at small interatomic distance, increases and becomes positive at larger interatomic distance and then would gradually decrease to zero when the wave function overlapping becomes negligible. This is the behavior of the single-peak Bethe-Slater curve, which describes the typical electron-electron exchange interaction as a function of the ratio interatomic spacing $r_{ab}$ and the radius of unfilled $d$ shell $r_d$ [45]. From this perspective, the double peak of $j_1$ and the single peak of $j_2$ can be understood.

From the values of $j_{ij}$ we may calculate the corresponding $T_C^{MFA}$ using Eq. (3), neglecting contributions from $j_3$ and above to avoid introducing more errors. The results are plotted in Fig. 6b as a function of the lattice constant. They are consistent with the experimental $T_C$ of BCC iron, which is known to be insensitive to pressure up to the -phase transition point at 1.75 GPa, [46] (i.e. from 2.856 Å to 2.8665 Å). They also agree very well with the GGA calculations of Kormann *et al.* [40], who showed that $T_C$ was not sensitive to pressure in the range 2.8 to 2.9 Å. Within the same range, the LDA results of [19] show that $T_C$ is not insensitive to, but decreases linearly with increasing lattice constant, similar to the non-convergent result that we could have obtained if we had included up to $j_8$ in our calculations.

**IV. Summary and Conclusion**

We may summarize our results as follows. According its defination [45], $J_{ij}$ should be dependent on all the on-site and inter-site interactions. All the factor affect these interactions, such as temperature, pressure, doping and etc., should affect the $J_{ij}$. However, it is difficult to give a unambivalent characterization between these factors and $J_{ij}$, not to mention getting mathematical formula for them. For example, our $J_{ij}$ data (Fig. 7a) show very complex pressure dependence. Here, we present a phenomenological description of the pressure dependence of $J_{ij}$ in BCC iron as following: when lattice lager than 2.94 Å, where on-site interaction play important role, the pressure dependences of $J_1$ and $J_2$ obey Bethe-Slater curver; for lower than 2.94 Å, both on-site and inter-site interactions are important, they compete and finally give a equiliubrium state. The subtle on-site and inter-site interactions is the base of complex pressure dependence. Just owing to this, iron has intriguing properties, such as Invar alloy, stainless steel and so on. The issue contributed to the magnetic properties related to $J_{ij}$ are very important both for fundamental science and technical application. With our *ab initio* magnetic moment data and spin-spin interatomic exchange interaction data, furthur study employed SLD will try to simulate more real materials with mesoscropic defects under room or high temperature.

In conclusion, employed LMTO green's function method, we have calculated the pressure dependence of exchange interactions $J_{ij}$ in BCC iron in a large volume extent. The results show rather complex pressure dependence of exchange interactions $J_{ij}$. This is consistent with the fact of pressure-induced decreasing of the ferromagnetic stability in BCC iron. Forthermore, we discussed the electronic structure changes due to pressure based on the FP-LAPW calculations. Finally, we suggest that introduce $J_{ij}$ to construct magnetic atomic potential is neccesary for SLD of iron with defect or dislocation.

**Acknowledgement**

This work was supported by grants from the Research Grants Council of the Hong Kong Special Administrative Region (Project: 530507 and 532008 ) and the Hong Kong Polytechnic University (Project Nos.: B-Q07M, B-Q14J and H-ZF20). We would like to thank Derek A. Stewart for providing the LMTO-GF code and for useful discussion.

**Reference:**

[1] H. C. Herper, E. Hoffmann *et al.*, Phys. Rev. B **60**, 3839 (1999).

[2] A. V. Ruban, P. A. Korzhavyi *et al.*, Phys. Rev. B **77**, 094436 (2008).

[3] D. J. Dever, J. Appl. Phys. **43**, 3293 (1972).

[4] H. Hasegawa, and D. G. Pettifor, Phys. Rev. Lett. **50**, 130 (1983).

[5] A. B. Sivak, V. A. Romanov *et al.*, in International Conference on Electron Microscopy and Multiscale Modeling, edited by A. S. Avilov *et al.* (Amer Inst Physics, Moscow, RUSSIA, 2007), pp. 118.

[6] M. L. Jenkins, Z. Yao *et al.*, J. Nucl. Mater. **389**, 197 (2009).

[7] S. Chiesa, P. M. Derlet *et al.*, Phys. Rev. B **79**, 214109 (2009).

[8] M. Y. Lavrentiev, R. Drautz *et al.*, Phys. Rev. B **75**, 014208 (2007).

[9] M. Y. Lavrentiev, S. L. Dudarev *et al.*, in 13th International Conference on Fusion Reactor Materials (ICFRM-13) (Elsevier Science Bv, Nice, FRANCE, 2007), pp. 22.

[10] G. Kresse, and J. Furthmuller, Comp. Mater. Sci. **6**, 15 (1996).

[11] V. P. Antropov, M. I. Katsnelson *et al.*, Phys. Rev. B **54**, 1019 (1996).

[12] D. Nguyen-Manh, A. P. Horsfield *et al.*, Phys. Rev. B **73**, 020101 (2006).

[13] P.-W. Ma, C. H. Woo *et al.*, Phys. Rev. B **78**, 024434 (2008).

[14] P. W. Ma, and C. H. Woo, Phys. Rev. E **79**, 046703 (2009).

[15] P.-W. Ma, C. H. Woo *et al.*, Philosophical Magazine **89**, 2921 (2009).

[16] O. N. Mryasov, A. J. Freeman *et al.*, J. Appl. Phys. **79**, 4805 (1995).

[17] S. Frota-Pessôa, R. B. Muniz *et al.*, Phys. Rev. B **62**, 5293 (2000).

[18] M. Pajda, J. Kudrnovský *et al.*, Phys. Rev. B **64**, 174402 (2001).

[19] S. Morán, C. Ederer *et al.*, Phys. Rev. B **67**, 012407 (2003).

[20] L. Stixrude, R. E. Cohen *et al.*, Phys. Rev. B **50**, 6442 (1994).

[21] A. Dewaele, P. Loubeyre *et al.*, Phys. Rev. Lett. **97**, 215504 (2006).

[22] O. K. Andersen, and O. Jepsen, Phys. Rev. Lett. **53**, 2571 (1984).

[23] P. Blaha, K. Schwarz *et al.*, Program for calculating crystal properties, WIEN2k (2001), Vienna University of Technology (ISBN 3-9501031-1-2).

[24] V. Iota, J.-H. P. Klepeis *et al.*, Appl. Phys. Lett. **90**, 042505 (2007).

[25] M. Ekman, B. Sadigh *et al.*, Phys. Rev. B **58**, 5296 (1998).

[26] M. van Schilfgaarde, and V. P. Antropov, J. Appl. Phys. **85**, 4827 (1999).

[27]     J. P. Perdew, K. Burke *et al.*, Phys. Rev. Lett. **77**, 3865 (1996).

[28]     J. P. Perdew, and A. Zunger, Phys. Rev. B **23**, 5048 (1981).

[29]     U. v. Barth, and L. Hedin, Journal of Physics C: Solid State Physics **5**, 1629 (1972).

[30]     C. S. Wang, B. M. Klein *et al.*, Phys. Rev. Lett. **54**, 1852 (1985).

[31]     M. Wuttig, and X. Liu, *Ultrathin Metal Films: Magnetic and Structural Properties* (Springer-Verlag, Berlin, 2004), Vol. Chap. 5.1.4. p. 138.

[32]     J. Mathon, Phys. Rev. B **27**, 1916 (1983).

[33]     O. Gunnarsson, O. Jepsen *et al.*, Phys. Rev. B **27**, 7144 (1983).

[34]     http://en.wikipedia.org/wiki/Exchange_interaction.

[35]     A. I. Liechtenstein, M. I. Katsnelson *et al.*, Journal of Physics F: Metal Physics **14**, L125 (1984).

[36]     F. Birch, Phys. Rev. **71**, 809 (1947).

[37]     P. D. a. N. S.-P. R. Kohlhaas, Z. angew. Phys. **23** 245 (1967).

[38]     M. Acet, H. Zähres *et al.*, Phys. Rev. B **49**, 6012 (1994).

[39]     J. Lazewski, P. Piekarz *et al.*, Phys. Rev. B **74**, 174304 (2006).

[40]     F. Kormann, A. Dick *et al.*, Phys. Rev. B **79**, 184406 (2009).

[41]     P. Richard, T. Sato *et al.*, Phys. Rev. Lett. **102**, 047003 (2009).

[42]     P. Söderlind, R. Ahuja *et al.*, Phys. Rev. B **50**, 5918 (1994).

[43]     A. K. McMahan, and R. C. Albers, Phys. Rev. Lett. **49**, 1198 (1982).

[44]     T. E. Jones, M. E. Eberhart *et al.*, Phys. Rev. Lett. **100**, 017208 (2008).

[45]     D. Jiles, *Introduction to Magnetism and Magnetic Materials* (Chapman & Hall, London, 1998).

[46]     J. M. Leger, C. Loriers-Susse *et al.*, Phys. Rev. B **6**, 4250 (1972).

**Tables**

Table I. By using LMTO-GF, the value of $j_{ij}$ (mRy) and the estimated $T_C^{MFA}$ for BCC iron are calculated at the experimental lattice constant (2.8665Å) as a function of the order of neighbour. Values from other works are also presented for comparison.

| $j_{ij}$ | GGA | LDA | Ref. [16][a] | Ref. [17][b] | Ref. [18][c] |
|---|---|---|---|---|---|
| $j_1$ | 1.218 | 1.235 | 1.2 | 1.24 | 1.432 |
| $j_2$ | 1.080 | 0.799 | 0.646 | 0.646 | 0.815 |
| $j_3$ | -0.042 | -0.009 | -0.030 | 0.007 | -0.015 |
| $j_4$ | -0.185 | -0.128 | -0.100 | -0.108 | -0.126 |
| $j_5$ | -0.117 | -0.093 | -0.068 | -0.071 | -0.146 |
| $j_6$ | 0.061 | 0.044 | 0.042 | 0.035 | 0.062 |
| $j_7$ | -0.013 | 0.001 | -0.001 | 0.002 | 0.001 |
| $j_8$ | 0.017 | 0.018 | 0.014 | 0.014 | 0.015 |
| $j_0$ | 12.57 | 12.41 | 11.03 | 12.38 | 13.58 |
| $T_C^{MFA}$ | 1323K | 1305K | 1160K | 1302K | 1428K |

[a]LDA; potential: von Barth and Hedin.

[b]LDA; potential: not mentioned.

[c]LDA; potential: Vosko-Wilk-Nusair.

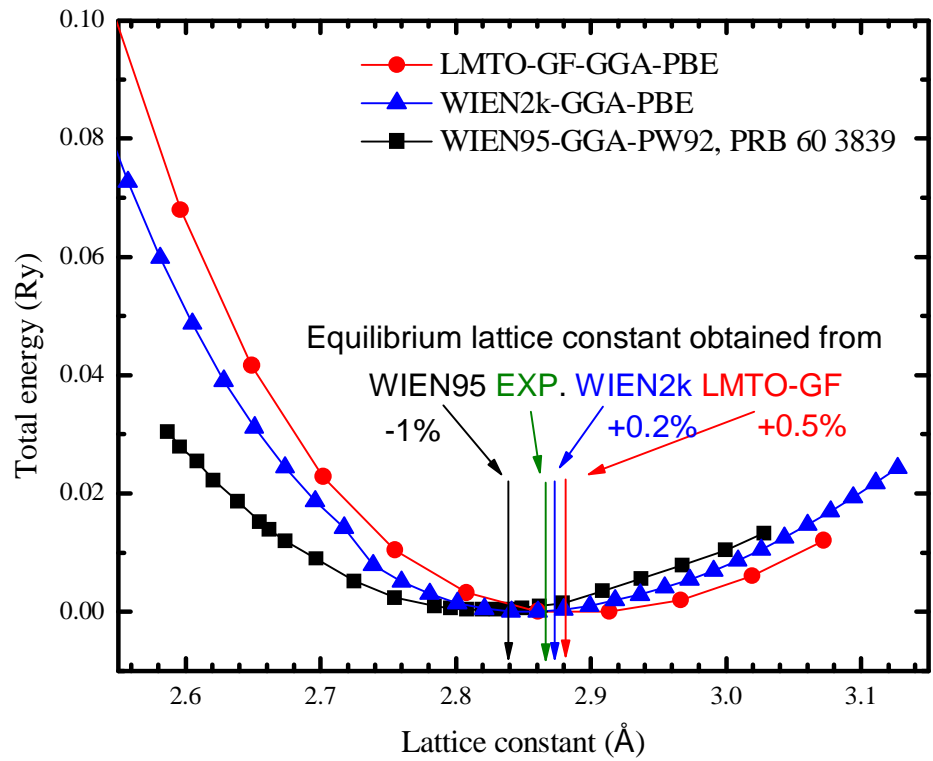

Fig. 1 (Color online). Calculated total energy as a function of lattice constant.

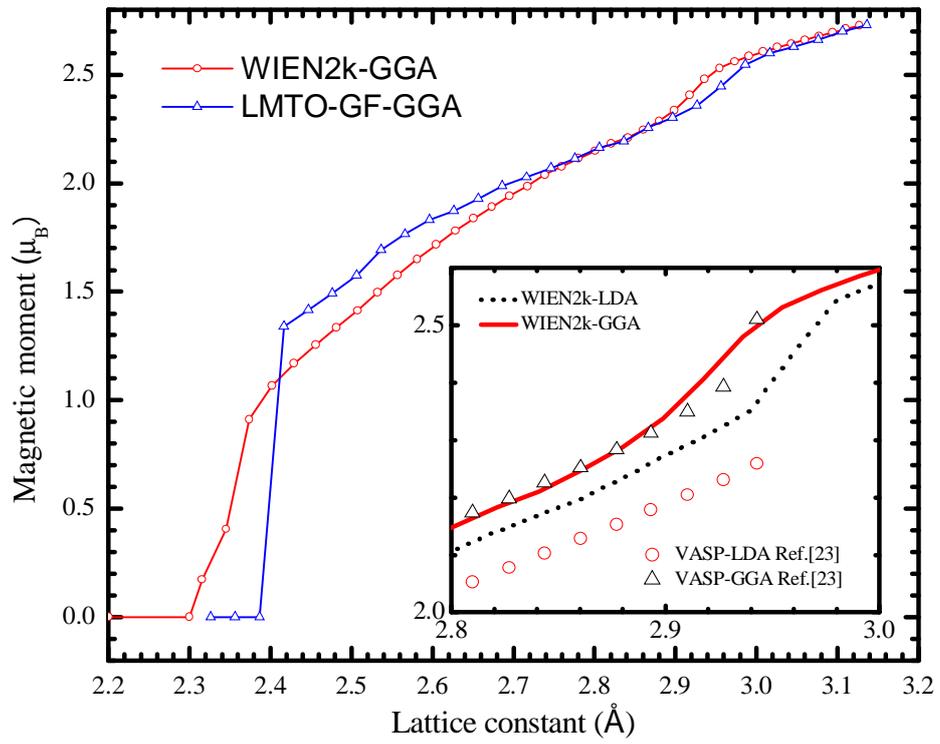

Fig. 2 (Color online). Calculated MM as a function of lattice constant.

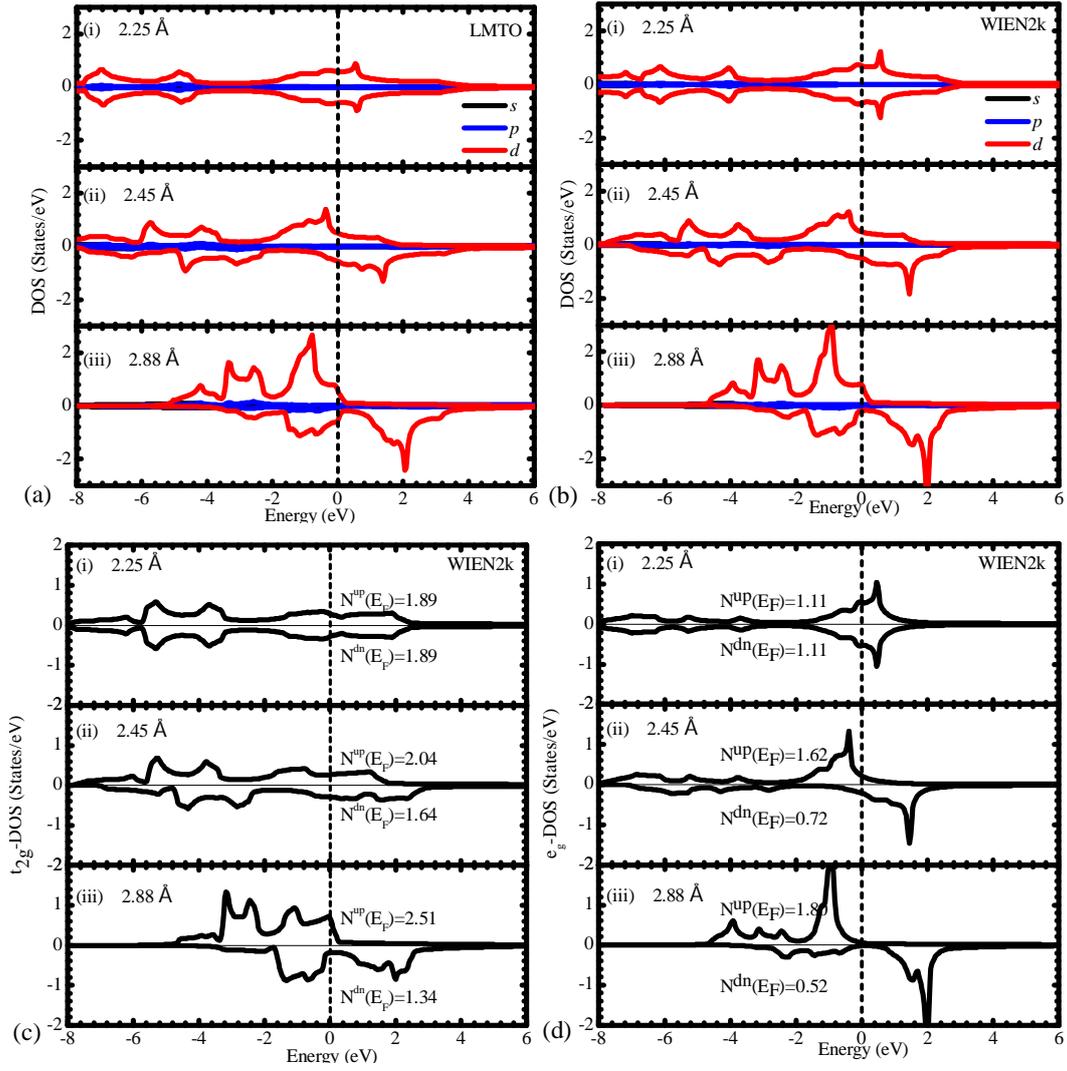

Fig. 3 (Color online). The partial DOS of BCC Fe using LMTO (a) and WIEN2k (b) methods for different lattice constants: 2.25 (i), 2.45 (ii) and 2.88 Å (iii). The $t_{2g}$ and $e_g$ 3d-DOS are shown in (c) and (d) pannels, respectively.

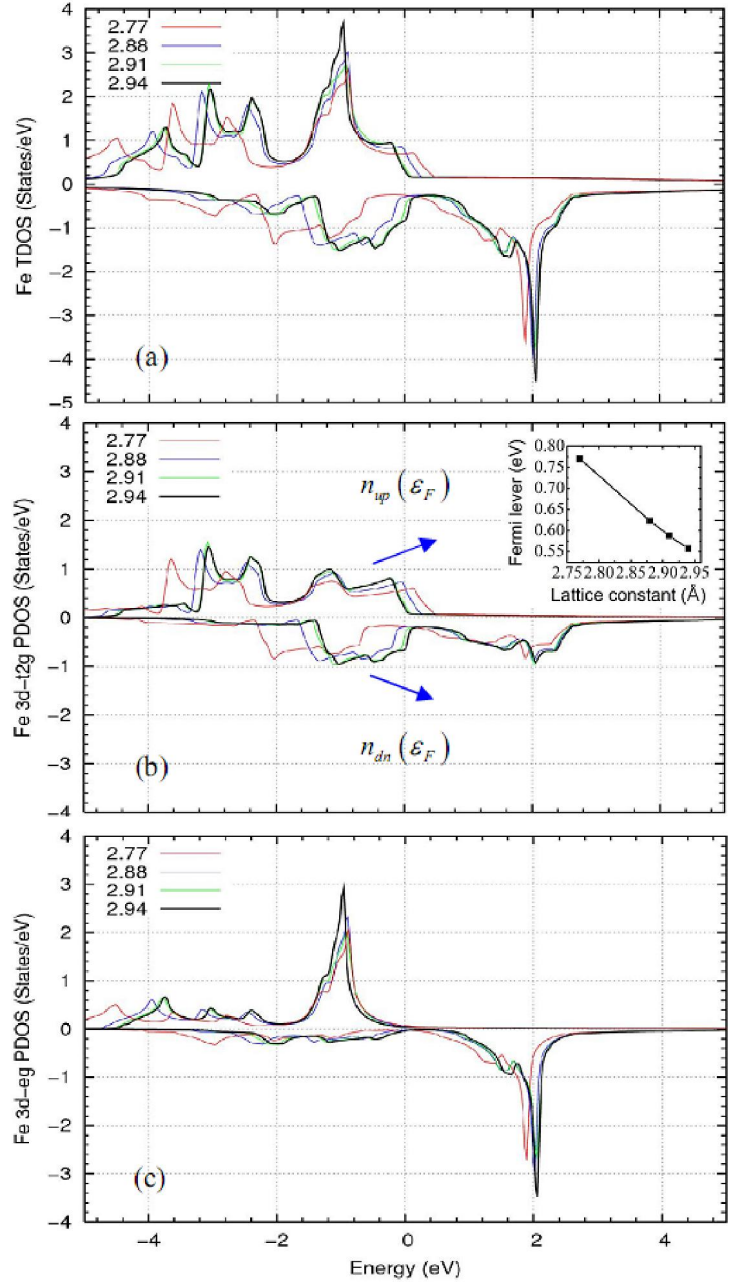

Fig. 4 (Color online). Total (a) and partial 3d-DOS (b, c) of BCC Fe for different lattice constants. The origin is set to its Fermi level for each lattice constant as shown in the insert of panel (b). The arrows show the change trend of DOS at Fermi level for both majority and minority states as decreasing the lattice constant.

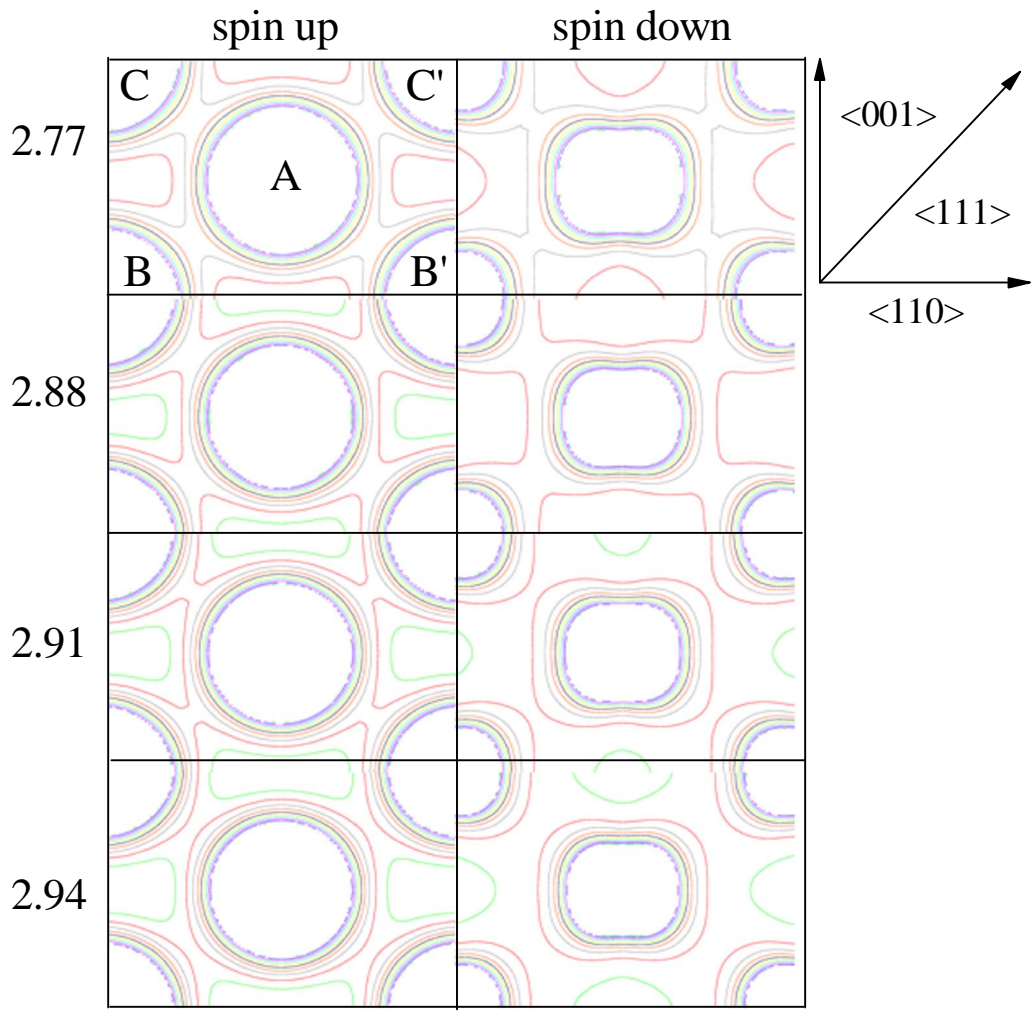

Fig. 5 (Color online). Charge density on (110) plane in BCC Fe for both majority and minority states.

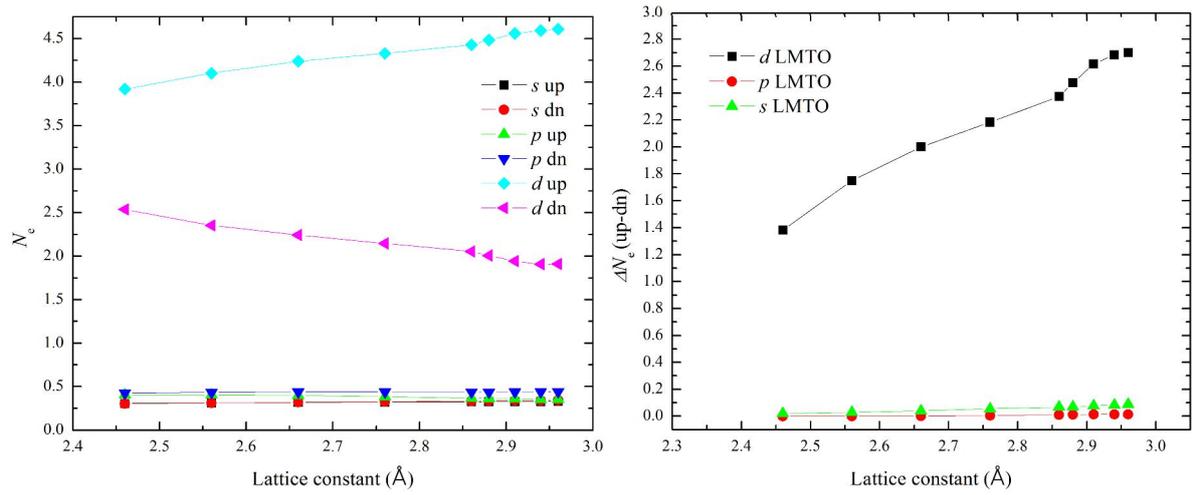

Fig. 6 (Color online). The occupancy number of electrons on *s*, *p*, and *d* orbitals for both majority and minority states (a), and the correspending magnetic moment compents (b) as a function of lattice constant. Here, date are obtained using LMTO method.

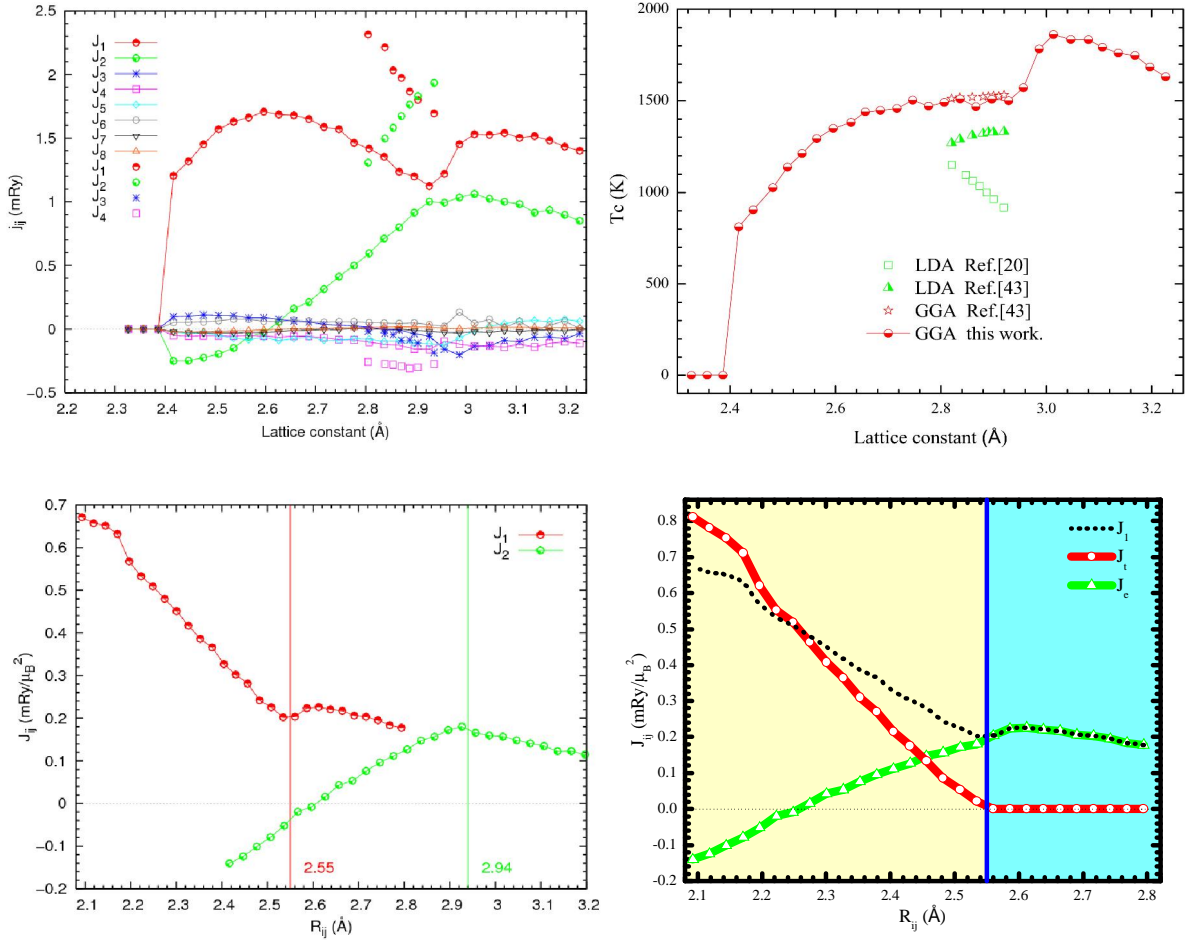

Fig. 7 (Color online). (a) The exchange interactions $j_{ij}$ from the 1$^{st}$ to the 8$^{th}$ neighbour as a function of lattice constant, where the point data are taken from Ref. [19]. (b) The Curie temperature is calculated according to mean field approximation using only $j_1$ and $j_2$. (c) The $J_{ij}$ from $J_1$ to $J_2$ as a function of R$_{ij}$ the distance between atoms $i$ and $j$. Here, R$_1=\frac{\sqrt{3}}{2}a$ and R$_2=a$ are the first and second neighbour distance, where $a$ is the lattice constant. The vertical lines show the location of lattice constant 2.94 Å for R$_1$ and R$_2$ respectively, near which the $J_1$ has a local minima while $J_2$ has a maximum.

22